\documentclass[12pt]{article}
\pdfoutput=1

\usepackage{cite}
\usepackage{mathtools}
\usepackage{amsfonts}
\usepackage{amssymb}
\usepackage{authblk}
\usepackage{bbold}
\usepackage{caption}
\usepackage{heck}
\usepackage[margin=1in]{geometry}
\usepackage{xcolor}
\usepackage{epsfig}
\usepackage{epstopdf}
\usepackage{graphicx}
\usepackage{siunitx}
\usepackage{hyperref}
\usepackage[nameinlink,capitalize]{cleveref}

\DeclarePairedDelimiter\abs{\lvert}{\rvert}
\DeclarePairedDelimiter\bra{\langle}{\rvert}
\DeclarePairedDelimiter\ket{\lvert}{\rangle}
\newcommand*{\braket}[2]{\ensuremath\langle#1\vert#2\rangle}
\newcommand*{\braopket}[3]{\ensuremath\langle#1\vert#2\vert#3\rangle}

\let\originalleft\left
\let\originalright\right
\renewcommand*{\left}{\mathopen{}\mathclose\bgroup\originalleft}
\renewcommand*{\right}{\aftergroup\egroup\originalright}

\DeclareMathOperator{\Tr}{Tr}

\DeclareMathOperator{\Var}{Var}
\newcommand*{\diff}{\mathop{}\!\mathrm{d}}

\newcommand*{\cD}{\ensuremath\mathcal{D}}
\newcommand*{\cH}{\ensuremath\mathcal{H}}
\newcommand*{\cL}{\ensuremath\mathcal{L}}
\newcommand*{\cM}{\ensuremath\mathcal{M}}
\newcommand*{\cN}{\ensuremath\mathcal{N}}
\newcommand*{\cO}{\ensuremath\mathcal{O}}

\begin{document}

\date{}

\title{Statistical coupling constants \\[4mm] from hidden sector entanglement}


\institution{PENN}{\centerline{${}^{1}$Department of Physics and Astronomy, University of Pennsylvania, Philadelphia, PA 19104, USA}}
\institution{BRUSS}{\centerline{${}^{2}$Theoretische Natuurkunde, Vrije Universiteit Brussel and}}
\institution{BRUSScont}{\centerline{International Solvay Institutes, Pleinlaan 2, B-1050 Brussels, Belgium}}

\authors{Vijay Balasubramanian\worksat{\PENN, \BRUSS}\footnote{e-mail: \texttt{vijay@physics.upenn.edu}},
Jonathan J. Heckman\worksat{\PENN}\footnote{e-mail: \texttt{jheckman@sas.upenn.edu}}, \\[4mm] Elliot Lipeles\worksat{\PENN}\footnote{e-mail: \texttt{lipeles@hep.upenn.edu}}, and Andrew P. Turner\worksat{\PENN}\footnote{e-mail: \texttt{turnerap@sas.upenn.edu}}}



\abstract{
String theory predicts that the couplings of Nature descend from dynamical
fields. All known string-motivated particle physics models also come with a
wide range of possible extra sectors. It is common to posit that such moduli
are frozen to a background value, and that extra sectors can be nearly
completely decoupled. Performing a partial trace over all sectors other than
the visible sector generically puts the visible sector in a mixed state,
with coupling constants drawn from a quantum statistical ensemble. An
observable consequence of this entanglement between visible and extra sectors
is that the reported values of couplings will appear to have an irreducible
variance. Including this variance in fits to experimental data gives an important
additional parameter that can be used to distinguish this scenario from
the case where couplings are treated as fixed parameters.
There is a consequent interplay between energy range and
precision of an experiment that allows an extended reach for new physics.
}

\maketitle

\section{Introduction}

The coupling constants of Nature are not truly ``constant.'' This, at least,
is what string theory predicts since such parameters descend from background
values of moduli fields, the low energy remnants of higher-dimensional quantum gravity in our
4D world. Parameters such as the fine structure constant or the top quark
Yukawa coupling are better viewed as dynamical---though perhaps
heavy---fields.

Indeed, one feature of all known string constructions is that beyond the
visible sector there are many additional degrees of freedom. These include
moduli (see~\cite{Coughlan:1983ci, Banks:1993en, Kane:2015jia}), as well as
large numbers of hidden sectors (see~\cite{Denef:2008wq}) which may only
weakly interact with the visible sector, typically through the mediation of
the moduli fields.
These possibilities pose a challenge in constructing phenomenologically viable
UV complete models, but also present an opportunity to access string-motivated
signatures of physics beyond the Standard Model.

Here, we observe that moduli fields that interact with hidden sectors will
necessarily be entangled with them. If so, following the standard rules of
quantum mechanics, we should trace out the hidden sector fields, thus deriving
an effective description of the moduli as being in a mixed quantum state. In
other words, the visible universe will be described by a quantum statistical
ensemble over couplings. We explain how to derive this ensemble, and describe
scenarios where there will be a measurable effect. In comparing with experiment,
this additional irreducible source of variance will generically produce a different
fit to the available data. This provides a way to distinguish this scenario from
the case where the couplings are treated as fixed parameters.



\section{The Visible Sector and Beyond}

Observationally, the visible sector is constructed from all the degrees of
freedom in the Standard Model, including 4D gravity. Observables of this
theory are often couched in terms of the particle content and possible
interaction terms, as governed by the coupling constants of a low energy
effective field theory. Letting $\cH_\text{vis}$ denote the Hilbert
space of states for the visible sector, there is a whole family of possible
ground states $\ket{\{\lambda\}} \in \cH_\text{vis}$ labelled by
the couplings of the theory.

The general message from string theory is that continuous couplings are really
background values for dynamical fields. This motivates promoting these
couplings to spacetime-dependent parameters $\{\lambda(x)\}$, and
correspondingly time dependent states $\ket{\{\lambda(\vec{x},t)\}} \in
\cH_\text{vis}$. In fact, to capture the full effects of this and other
possible string-motivated degrees of freedom, it is more appropriate to
enlarge the associated Hilbert space. In what follows, we shall assume that
there is an approximate factorization of the full Hilbert space as:
\begin{equation}
    \cH_\text{full} \approx \cH_\text{vis} \otimes \cH_{\overline{\text{vis}}}\,,
\end{equation}
where the factor $\cH_{\overline{\text{vis}}}$ denotes everything ``other than
the visible sector.''

\begin{figure}[t!]
    \centering

    \includegraphics[trim={0 0cm 0 0},clip,scale=0.35]{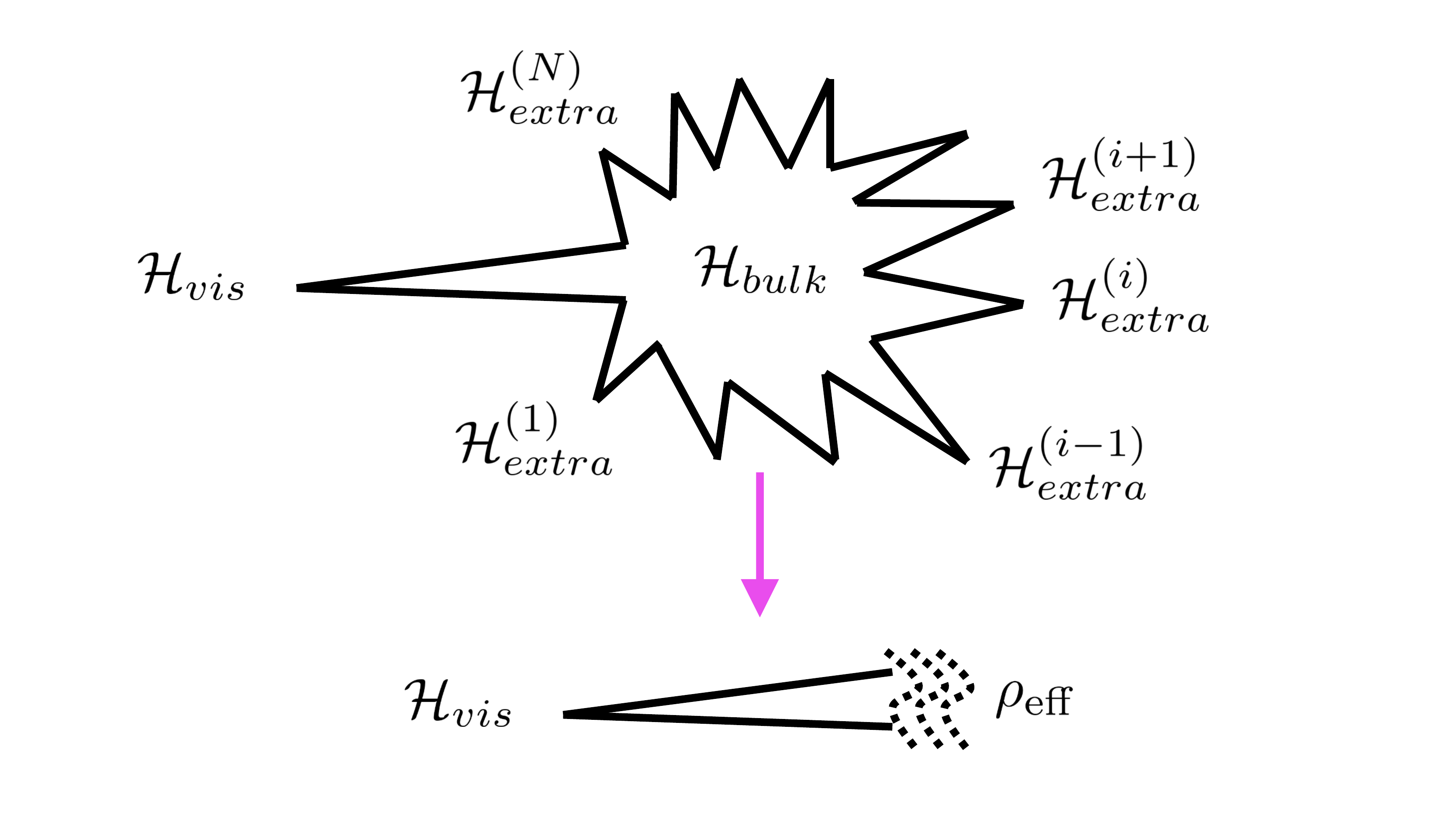}

    \caption{Depiction of the many QFT sectors of a string compactification.
    Performing a partial trace over the complement of the visible sector yields an
    effective density matrix for the visible sector couplings.}
    \label{fig:VisBulkExtraV3}
\end{figure}

This sort of factorization is  well-motivated in the context of string
constructions (see, e.g.,~\cite{Heckman:2010bq} for an early review of some
F-theory examples) since the Standard Model is typically localized on a
subspace of the full higher-dimensional system, and many extra sectors are
sequestered at other locations of the extra-dimensional geometry. There can
still be mixing between these sectors through bulk modes such as closed string
moduli, which permeate the extra-dimensional geometry, and there are
Swampland arguments that some such mixing is irreducible \cite{Heckman:2019bzm}.
In such cases, additional structure is present, and we can write:
\begin{equation}
    \cH_{\overline{\text{vis}}} =
    \cH_\text{bulk} \otimes \cH_\text{extra}^{(1)} \otimes \dots \otimes \cH_\text{extra}^{(N)}\,,
\end{equation}
in the obvious notation (see \cref{fig:VisBulkExtraV3}).

There are clearly a wide range of possibilities for the dynamics in
$\cH_{\overline{\text{vis}}}$. If all the mass scales of non-visible sector
states are heavy, it is appropriate to use the general framework of
effective field theory, integrating out the heavy modes. In the
visible sector, this will be encoded in a particular structure for a low
energy effective action for visible sector states.

It can also happen, however, that the states of the extra sector are
 light, perhaps comparable with their visible sector counterparts. In such
cases, integrating out these degrees of freedom would result in a non-local
effective action in the visible sector. One (practically quite
cumbersome) way to track the effects of such extra sectors is via non-analytic
behavior in various correlation functions/scattering amplitudes.

We will purse an alternative approach, by treating the visible sector as an
open system. Letting $\rho_\text{full}$ denote the density matrix for the full
system, we obtain a reduced density matrix $\rho_\text{vis} \in \cH_\text{vis}
\otimes \cH_\text{vis}^*$ for just the visible sector by performing a partial
trace over the complement:
\begin{equation}
    \rho_\text{vis} = \Tr_{\cH_{\overline{\text{vis}}}} \rho_\text{full}\,.
\end{equation}
So even if $\rho_\text{full}$ is a pure state, such as the ground state of the
full system, the reduced density matrix will typically be a mixed state.  We will focus on the part of this mixed state
that involves the couplings.  Note also that since there could
be some unknown dynamics in the extra sector, we should
treat $\rho_\text{vis}(t)$ as a time-dependent state. From
the perspective of string compactification, this really is a geometric
entanglement, since it involves tracing over all regions of the
geometry other than where the visible sector is localized.
Some explicit examples of stringy brane systems where this sort of entanglement across different sectors
was considered include references \cite{Heckman:2020otd, Balasubramanian:2020ffd}.

On general grounds, the mixed state $\rho_\text{vis}$ will involve a sum over
possible spacetime configurations for the couplings $\{\lambda(\vec{x}, t)\}$,
which we summarize schematically as:
\begin{equation}
    \rho_\text{vis} = \sum_{\lambda, \lambda'} \rho_\text{vis}(\lambda, \lambda') \ket{\lambda} \bra{\lambda'}\,.
\end{equation}
Said differently, the partial trace over the complement of the visible sector
produces a quantum statistical ensemble over spacetime-dependent couplings.

\section{Example: Coupled Oscillators}
\label{sec:osc}

The general considerations are already clear in the simple example of a pair
of coupled oscillators with a Hamiltonian
\begin{equation}
    H = \frac{1}{2} \left(p_{x_1}^2 + p_{x_2}^2\right) + \frac{1}{2} \left(\omega_1^2 x_1^2 + \omega_2^2 x_2^2 + \lambda x_1 x_2\right)\,.
\end{equation}
We will think of $x_1$ as a modulus field and $x_2$ as a field in the hidden
sector. There may be additional interactions between $x_1$ and other visible
sector degrees of freedom, but we neglect them here because we are interested
in the effect on $x_1$ of tracing over $x_2$. The general case of coupling to
many hidden sector oscillators is treated in \cref{sec:many-osc}.

We can diagonalize this Hamiltonian in terms of the variables $y_1 = x_1
\cos\alpha - x_2 \sin\alpha$ and $y_2 = x_1 \sin\alpha + x_2 \cos\alpha$ with
the mixing angle specified by $\tan2\alpha = \lambda / (\omega_2^2 -
\omega_1^2)$. The Hamiltonian becomes
\begin{equation}
    H = \frac{1}{2} \left(p_{y_1}^2 + p_{y_2}^2\right)
    +
    \frac{1}{2} \left(\omega_1'^2 y_1^2 + \omega_2'^2 y_2^2\right)\,,
\end{equation}
where
$\omega_1'^2 = \omega_1^2 - (\lambda / 2) \tan\alpha$ and $\omega_2'^2 =
\omega_2^2 + (\lambda / 2) \tan\alpha$. The ground state is then a product of
Gaussians in $y_1$ and $y_2$.

Converting back to the original physical variables we find the ground state wavefunction
\begin{equation}
    \psi_0 = \frac{(\det A)^{1 / 4}}{\sqrt{\pi}}
    e^{-\frac{1}{4} \left(\frac{x_1^2}{\tau_1^2} + \frac{x_2^2}{\tau_2^2} + \frac{2 x_1 x_2}{g}\right)}\,, \quad A=
    \begin{pmatrix}
       \frac{1}{2 \tau_1^2} & 1 / 2 g \\
        1 / 2 g & \frac{1}{2 \tau_2^2}
    \end{pmatrix}
\end{equation}
where
\begin{equation}
    \frac{1}{2 \tau_1^2} = \omega_1' \cos^2\alpha + \omega_2' \sin^2\alpha\,, \quad \frac{1}{2 \tau_2^2} = \omega_2' \cos^2\alpha + \omega_1' \sin^2\alpha\,, \quad \frac{1}{2 g} = \left(\omega_2' - \omega_1'\right) \sin\alpha \, \cos\alpha\,.
\end{equation}
We see that when $g$ is finite, the modulus field $x_1$ and the hidden sector
field $x_2$ are entangled, i.e., their wavefunction does not factorize, and in
the $g \to \infty$ limit, the entanglement is weak. Tracing out $x_2$ gives a
density matrix on $x_1$: $\rho = \int \diff x_2 \; \psi_0(x_1, x_2) \,
\psi_0^*(x_1', x_2') = \Tr_{x_2}(\psi_0^* \psi_0)$.  We find that
\begin{equation}
    \rho(x_1, x_1') = \frac{\sqrt{\det A}}{\pi} \sqrt{2 \pi \tau_2^2}
    e^{-(x_1^2 + x_1'^2) / 4 \tau_1^2} e^{\tau_2^2 (x_1 + x_1')^2 / 8 g^2}\,,
\end{equation}
from which we can work out  the variance of the modulus $x_1$ when $x_2$
is not observed:
\begin{equation}
    \Var(x_1) = \tau_1^2 \left(1 - \frac{\tau_1^2 \tau_2^2}{g^2}\right)^{-1}
    \xrightarrow{g\to \infty}
    \tau_1^2 \left(1 + \frac{\tau_1^2 \tau_2^2}{g^2}\right)\,,
\end{equation}
where the equation on the right is in the limit of weak entanglement.

Several qualitative features are clear from these results. Suppose $\lambda =
0$, so that the modulus does not interact with the hidden sector. Then the
mixing angle is $\alpha = 0$, $\omega_1' = \omega_1$, $g = \infty$, $\tau_1^2
= 1 / 2 \omega_1$, and the variance of $x_1$ is simply fixed by the steepness
of its potential, $\Var(x_1) \propto 1 / \omega_1$, so that a shallow
potential (small $\omega_1^2$) leads to large fluctuations. We want to show
that even if the potential for the modulus is steep (large $\omega_1^2$), the
variance of $x_1$ can still be large because of the interaction with the
hidden sector. Notice that if we tune $\omega_1$, $\omega_2$, and $\lambda$ so
that $g \to \tau_1 \tau_2$ at fixed $\tau_1$ and $\tau_2$, the variance of
$x_1$ necessarily becomes large.  Mechanistically, this is because even if
both $x_1$ and $x_2$ are stiff directions, their mixing can generate a shallow
direction in the combined potential. Large quantum
fluctuations along this valley contribute to the variance of $x_1$.
Here, we are treating the same phenomenon in the language of open quantum
systems, with $x_1$ entangled with a hidden bath. If there are $N$ independent
hidden sector oscillators mixing with with the modulus in this way, even weak
mixing can generate a large effect, because the variance will be enhanced as a function of
$N$.




\section{Entangled Moduli in Field Theory}

Let us now turn to the case of a quantum field theory engineered via string theory.
A common situation is that we get a Lagrangian which depends on some coupling constants.
In the visible sector Lagrangian, this appears
through a term of the form:
\begin{equation}
    \cL_\text{v} \supset \lambda_\text{v} \cO_{\text{v}}\,,
\end{equation}
where $\lambda_{\text{v}}$ is a visible sector coupling, and $\mathcal{O}_{\text{v}}$ is a
visible sector operator. We promote $\lambda_{\text{v}}$ to a
dynamical field $\phi$ with a ``decay constant''
$f_\text{v}$, performing the substitution:
\begin{equation}
    \lambda_\text{v} \mapsto \lambda_\text{v} + \frac{\phi}{f_\text{v}}\,.
\end{equation}
Doing so motivates us to consider an enlarged Hilbert space that includes the
modulus as well as its possible couplings to other sectors. For example, this modulus
can appear equally well as a coupling in an extra sector, so we generically expect mixing terms
involving visible and hidden sector operators $\cO_\text{v}$ and
$\cO_\text{h}$:
\begin{equation}
    \label{eq:closeddecayconst}
    \cL_\text{mix} = \left(\lambda_\text{v} + \frac{\phi}{f_\text{v}}\right) \cO_\text{v}
    + \left(\lambda_\text{h} + \frac{\phi}{f_\text{h}}\right) \cO_\text{h}\,.
\end{equation}

There is no shortage of examples coming from string theory. For example, if we interpret $\phi$ as a
closed string modulus, the associated decay constant might be Planck or GUT scale, some axion models
have lower decay constant scales, while in some models where $\phi$ is instead an open string modulus,
the scale could be far closer to the  TeV range \cite{Heckman:2015kqk}, if the mass scales are correlated
with supersymmetry breaking.

In most cases, one typically assumes there is some potential that stabilizes the value of
$\phi$ at zero. This potential, as well as the various mass scales, decay constants, and number of
extra sectors introduces a large number of possibilities, and with it a seemingly
endless variety of possible signatures.

We seek a more model independent way to characterize the range of possible signatures. To this end,
we can perform a partial trace on $\rho_\text{full}$ over the complement of $\cH_\text{vis}$ states, and thus
obtain a mixed state for the visible sector density matrix, with a distribution of
possible coupling constants.

Let us now illustrate how the density matrix for couplings comes about.
The ground state is constructed by the Euclidean path integral as
\begin{equation}
    \label{eq:gndstate}
    \langle \Phi , \Phi_h \vert \Psi \rangle
    =
    \int^{\Phi, \Phi_\text{h}, t = 0}_{t = -\infty}
    \cD \phi \,
    \cD \phi_\text{h} \;
    e^{-S}\,,
\end{equation}
where we have suppressed indices on the visible and hidden sector fields and
indicated schematically that the integral sums over all configurations from $t
= -\infty$ to $t = 0$ with the boundary condition that $\phi(0^-) = \Phi$ and
$\phi_\text{h}(0^-) = \Phi_\text{h}$, with capital variables serving to emphasize that
these are the spatial profiles of some field at a fixed time. The density matrix for the system
\begin{equation}
    \rho = \ket{\Psi}\bra{\Psi}\,,
\end{equation}
is constructed by multiplying the state vector in \cref{eq:gndstate} by its
conjugate, computed by path integrating from $t = 0^+$ to $t = \infty$ with
the boundary condition $\phi(0^+) = \Phi'$ and $\phi_\text{h}(0^+) =
\Phi_\text{h}'$.

The reduced density matrix for the modulus field is obtained by tracing out
the hidden fields:
\begin{equation}
    \rho_\text{red} = \Tr_{\phi_\text{h}}(\ket{\Psi}\bra{\Psi})\,.
\end{equation}
In path integral language, this amounts to setting $\phi_\text{h}(0^-) =
\phi_\text{h}(0^+) = \Phi_\text{h}$ in the integral for the density matrix and
then integrating over $\phi_\text{h}$ for all times, including the $t=0$ boundary value $\Phi_h$.  As discussed in
\cite{Balasubramanian:2011wt}, this is equivalent to first integrating out the
hidden field $\phi_\text{h}$ completely to get a quantum effective action
$S_\text{eff}(\phi)$ for the modulus field, and then doing the construction of
the density matrix for $\phi$ with the exponential of the effective action
weighting the path sum.  From this point of view, the equal time correlation
function of operators $\cO$ constructed out of the modulus at $t = 0$ is
\begin{equation}
    \langle\cO \cO\rangle =
    \Tr(\cO \cO \rho_\text{red})
    = \int \cD \phi \,
    \cD \phi_\text{h} \; \cO \, \cO \,
    e^{-S}\,.
 \end{equation}
To arrive at the expression on the right hand side, we carry out the path integral for the wavefunctional ($\Psi(\Phi,\Phi_h) = \braket{\Phi, \Phi_h}{\Psi}$) and its conjugate ($\Psi^*(\Phi',\Phi_h') = \braket{\Psi}{\Phi', \Phi_h'}$) to find the density matrix as described above, then sew the boundary conditions across the $t=0$ for the hidden sector ($\Phi_h=\Phi_h'$) and integrate to find the reduced density matrix, and then finally multiply by $\cO\cO$, sew the boundary conditions for the modulus field $\phi$ across $t=0$ ($\Phi = \Phi'$) and integrate to take the trace.  Overall, this gives a path integral over the values of the fields at all times as shown.
The last expression shows the relation between the reduced density matrix formulation of
correlators of $\phi$ and the standard path integral for the same quantities.

We want to work out the density matrix for the modulus $\phi$ in the vacuum
state $\ket{0}$ of our field theory:
\begin{equation}
    \rho_\text{red}(\Phi, \Phi') =
    \int \cD\Phi_\text{h} \,
    \braket{\Phi', \Phi_\text{h}}{\Psi} \braket{\Psi}{\Phi, \Phi_\text{h}}
    =
    \cN
    \int \cD\Phi_\text{h} \; \Psi^*(\Phi',\Phi_h) \Psi(\Phi,\Phi_h)
    \,.
\end{equation}
Here $\ket{\Phi,\Phi_h}$ is a projector onto the $\Phi$, $\Phi_h$ field configuration.
We will consider a simple toy model that illustrates the general point, with a
quadratic Lagrangian written schematically as
\begin{equation}
    \label{eq:Lagrangian}
    \cL \sim
    \phi G \phi
    + \phi_\text{h}^i G_{i j} \phi_\text{h}^j
    + \phi \lambda_j \phi_\text{h}^j\,.
\end{equation}
where we have integrated the action by parts and dropped boundary terms to
write the local Lagrangian density in terms of quadratic differential
operators $G$ and $G_{i j}$  (e.g., $G \sim \Box + m^2$) and the coupling
$\lambda_j$.  The ground state wavefunctional is then
\begin{equation}
    \Psi(\Phi, \Phi_\text{h}^i) = \int^{\Phi, \Phi_\text{h}^i, t=0}_{t=-\infty} \cD\phi \, \cD\phi_\text{h}^i \; e^{-\int d^{D} x \, \cL}\,.
\end{equation}
Since the action is quadratic, similarly to the harmonic oscillator example
that we described above, the ground state wavefunctional will also be
quadratic
\begin{equation}
    \label{eq:wavefunction}
    \Psi(\Phi, \Phi_\text{h}^i) = \sqrt{\cN}
    e^{-\frac{1}{2} \int \diff^{D - 1} x \;
    \Phi \Omega \Phi + \Phi_\text{h}^i \Omega_{i j} \Phi_\text{h}^j
    + \Phi \epsilon_j \Phi_\text{h}^j}\,,
\end{equation}
where $\cN$ normalizes $\Psi$. We are being schematic here---strictly
speaking, the exponent in the wavefunctional will be a bi-local integral, and
$\Omega$, $\Omega_{i j}$ and $\epsilon_j$ will be complicated functions of
$G$, $G_{i j}$, and $\lambda_j$. We will simply be interested in the scaling
of these quantities as we intend this as a toy model.

The reduced density matrix is then:
\begin{equation}
    \rho_\text{red} = \cN
    \int \cD\Phi_\text{h}^i \; \Psi^*(\Phi',\Phi_h) \Psi(\Phi,\Phi_h)\,.
\end{equation}
This is a Gaussian integral, and can be evaluated explicitly to give
\begin{equation}
    \label{eq:RedModDensity}
    \rho_\text{red}
    =
    \cN e^{-\frac{1}{2}
    \int \diff^{D - 1} x \;
    [(\Phi - \Phi') (\Omega / 2) (\Phi - \Phi')
    +
    (\Phi + \Phi') \Omega_\text{eff} (\Phi + \Phi')]
    }\,,
\end{equation}
where
\begin{equation}
    \Omega_\text{eff}
    =
    \frac{\Omega}{2}
    -
    \frac{1}{8} \epsilon_i \Omega_{i j}^{-1} \epsilon_j\,.
\end{equation}
Finally, we are in a position to evaluate the variance of the modulus field
\begin{equation}
    \langle\Phi \Phi \rangle =
    \Tr(\rho_\text{red} \Phi \Phi)
    = \cN \int \cD\Phi \; \Phi \, \Phi \,
    e^{-\int \diff^{D - 1} x \; \Phi \Omega_\text{eff} \Phi}\,,
\end{equation}
where we set $\Phi = \Phi'$ in \cref{eq:RedModDensity} and then integrated
over $\phi$ to take the trace. Thus, the equal-time variance is
\begin{equation}
    \label{eq:var}
    \Var(\Phi) = \Omega_\text{eff}^{-1}\,.
\end{equation}
Again, we are being schematic.  More generally we are here really describing
the equal time correlation function at some separation, and when the separation
is small this correlator measures the variance in the field.

We want to know whether the variance \labelcref{eq:var} can be large. The basic
scale for the variance of $\phi$ is set by the $\Omega$ in the wavefunctional
\labelcref{eq:wavefunction}. To estimate the effect of the
coupling to the hidden sector, let us assume that all the quantities in the
wavefunctional have similar orders of magnitude $O(\Omega) \sim O(\Omega_{i j})
\sim O(\epsilon_j) \, \forall i,j$.  In fact, for concreteness let us take
$\Omega_{i j} \sim \Omega I_{N \times N}$ and $\epsilon_i \sim \epsilon
\Omega$. Then
\begin{equation}
    \Omega_\text{eff} \sim \frac{1}{8} (4 - \epsilon^2 N) \Omega\,. 
\end{equation}
We see that for any $\epsilon$ and $\Omega$, if the number of hidden sector fields $N$ is large, it can easily happen that $\Omega_\text{eff} \ll \Omega$, so that $\Var(\phi) \gg 1 / \Omega$. (Note that we must have $\epsilon^2 N < 4$ to maintain positive-definiteness of the coupling matrix.)  In other words, if there are many hidden sector fields, as there typically are in string theory, they can have a substantial effect on the variance of a heavy modulus field with which they are weakly coupled.  Alternatively, we can imagine that the hidden sector fields are lighter than the modulus, as there is nothing forbidding this. This means that $\abs{\Omega_{i j}} \ll \abs{\Omega}$, and can also lead to a small $\Omega_\text{eff}$ and thus a large variance for the modulus field.

As we will discuss below, in typical experiments the interactions occur at different times, so we are really interested in the unequal time correlators of $\phi$.  To study this in the language of the reduced density matrix, we must time-evolve it, a dynamics that is typically not Hamiltonian, but controlled rather by the Lindblad equation, unless the measurements are appropriately coarse-grained in time \cite{Agon:2014uxa}.

\section{Couplings and Correlators}

In the previous sections we emphasized that tracing over the extra sector states means that in general, the
visible sector actually probed by experiment is really in a mixed state, and consequently, that there is a
statistical ensemble of possible values for the coupling constants. Note that even at equal times this can lead to non-trivial spatial correlations for couplings. In practice, carrying out explicit calculations  in this setup is somewhat awkward because the very appearance of a wavefunction references a preferred time slicing  of our spacetime. If our eventual aim is to extract observables as obtained from a scattering experiment,  we should also seek out a treatment which is suitably Lorentz covariant. Again taking our cue from string theory where  such couplings descend from moduli fields, we know that the appropriate way to analyze such structures  is in terms of the Lorentz covariant correlation functions of the moduli fields. We can visualize this as breaking up the  spacetime into small four-dimensional ``pixels'' and assigning a particular value of the coupling in each such pixel. In the limit where the pixels are quite small, we expect the correlation function to assume a delta function approximation:
\begin{equation}
\langle \lambda(x) \lambda(x^{\prime}) \rangle \sim c M_\text{UV}^{-4} \delta^{4}(x - x^{\prime}),
\end{equation}
with $M_\text{UV}$ some UV mass scale, and $c$ a model dependent parameter.

Suppose now that we perform a scattering experiment involving visible sector states. The amplitude can be packaged as a correlation function of visible sector operators $\cO_\text{vis}^{(i)}$ evaluated in the mixed state $\rho_\text{vis}(t)$ obtained by tracing over
both bulk moduli and extra sector states:
\begin{equation}
    i \cM \sim \Tr_{\cH_\text{vis}}\left(\rho_\text{vis}(t)
    \cO_\text{vis}^{(1)} \dotsm \cO_\text{vis}^{(n)}\right)\,.
\end{equation}
One can first perform all visible sector correlation functions, and then perform a further evaluation of all correlators involving the couplings. This is valid to do in a decoupling limit of string theory, and is reminiscent of the procedure one adopts in disorder averaging, though the interpretation is somewhat different.\footnote{See references~\cite{Rothstein:2012hk, Green:2014xqa, Craig:2017ppp} for some applications of disorder averaging in particle physics and cosmology.}

When we report the value of a coupling constant, we are working backwards from the measured cross section to a corresponding scattering amplitude to extract the value of the coupling constants in our underlying theory. Let us call this reported value of the coupling $\lambda_\text{expt}$.
This of course comes with a central value as well as some variance. As one improves the precision of an experiment, one expects to reduce this variance.

But, as we have already seen, by treating the visible sector as an open
system, there is always an irreducible amount of variance we get just from
tracing over everything other than the visible sector. In fact, we can
estimate the impact of this just by comparing the values of scattering
amplitudes we get by treating $\lambda(\vec{x}, t)$ as a statistical
parameter. For example, if we have a specific model of physics beyond the
visible sector in mind, we can extract the two-point function for couplings
via
\begin{equation}
    \Tr_{\cH_\text{vis}}(\rho_\text{vis}(t)
    \lambda(x) \lambda(x')) = \Tr_{\cH_\text{full}}(\rho_\text{full}
    \lambda(x) \lambda(x'))\,,
\end{equation}
that is, by evaluating the two-point function in the full Hilbert space.

Such perfect knowledge of the extra sectors is typically unavailable. This
motivates seeking alternative ways to package the possible effect on visible
sector observables. Along these lines, we can think of an observer performing
an experiment at some energy scale $Q_\text{expt}$ as ``sampling'' from a
probability distribution of couplings. Each point in space and time gives a
unique sampled value.

Our observer works in a small spacetime volume of size $\Vol_\text{4D} \sim
Q_\text{expt}^{-4}$. In each such chunk, they can approximate the sampled
value of the coupling by a pure number, call it $\lambda_\text{expt}$. We
assume the leading order variation of the probability distribution is
governed by a Gaussian centered on $\lambda _\text{expt} = \lambda_0$ (as happens in
the ground state of the harmonic oscillator):
\begin{equation}
    P(\lambda_\text{expt}) \sim \exp\left(-\frac{(\lambda_\text{expt} - \lambda_0)^2}{2 (\Delta\lambda_\text{expt}^2)} + \dotsb\right)\,,
\end{equation}
with variance:
\begin{equation}
    \Delta\lambda_\text{expt}^2 \sim \left(\Vol_\text{4D} \times
    M_*^{4}\right)^{-1} = Q_\text{expt}^4 / M_*^4\,.
\end{equation}
Here, $M_*$ is some characteristic mass scale that folds in all the information of the extra sectors.\footnote{Two powers of the mass scale come from the dimension of the field, and two powers come from the dimension of the decay constant $f$ in \cref{eq:closeddecayconst}.}  In actual scenarios, the specific details for how this scale is generated could be wildly different. None of this matters for this class of observables.  The volume dependence can be tracked by considering the integrals involved in computing a correlator. But, conceptually, we can think about it by imagining that in an experimental volume $\Vol_\text{4D} \sim
Q_\text{expt}^{-4}$, every appearance of a coupling in a process is averaged via the path sum over independent samples at a number of points that is proportional to this volume.   Then we can estimate that the standard deviation of the apparent coupling will decrease by a factor $1/\sqrt{\Vol_\text{4D}}$ and thus the variance will be suppressed by $1/\Vol_\text{4D}$.

Standard lore  holds that since we are at energies low compared to the scale at which the modulus fluctuates, we can treat the couplings as frozen, position-independent parameters. Indeed, if $Q_\text{expt}$ is far below
$M_*$, we have a sharply peaked Gaussian. This, however, only covers a subset of well-motivated possibilities, even when the modulus is heavy. Indeed, as seen in the example of a harmonic oscillator in \cref{sec:many-osc}, the ``extra sector'' could consist of many additional light degrees of
freedom. The general point is that if there are many extra sectors, and especially if some are at strong coupling, there is a general broadening of the associated distribution of couplings.

\section{Signatures}

We now ask whether it is possible to measure this effect in actual
experiments. A common way to look for extra sectors is to study apparent
violations of conservation laws, e.g., missing transverse energy signals. This
only covers some models. Examples include effects from soft radiation to an
extra sector. The point of the present approach is that even in the absence of
more direct signatures, it is still possible to look for potential effects
from such sectors.

A nonzero measured variance in the couplings will show up in processes that
scale with different powers of the couplings, and consequently the difference
$\Delta\lambda_\text{expt}^2 = \langle\lambda_\text{expt}^2\rangle -
\langle\lambda_\text{expt}\rangle^2$. Note that an effect suppressed by more
insertions of the coupling can easily be overcome by kinematic effects, namely
a bigger jump in the transition energies of the system. We leave an exhaustive
study of possible signatures for future work.

Our relation between effective mass scales and the distance scale being probed
extends the traditional ``reach'' of an experiment. If we observe a null
result at some energy scale $Q_\text{expt}$ and precision
$\Delta\lambda_\text{expt}$, then we get a mass scale limit:
\begin{equation}
    M_\text{limit} = \abs{\Delta\lambda_\text{expt}}^{-1 / 2} Q_\text{expt}\,.
\end{equation}

The best limits can be set either by having a very precise measurement, or
alternatively, going to much higher energy scales. We take as representative
examples atomic physics experiments and collider physics experiments:
\begin{equation}
    Q_\text{atomic} \sim \SI{10}{\eV}\,, \quad Q_\text{collider} \sim \SI{1}{\tera\eV}\,.
\end{equation}
The current precision of the fine structure constant is on the order of $\sim
10^{-10}$, and one can anticipate determining some couplings at the LHC at the
level of $\sim 10^{-2}$, as in~\cite{Farina:2016rws}. Plugging in for these
quantities, we see that depending on the coupling constant and the underlying
mass scales, we can set limits:
\begin{align*}
    M^\text{atomic}_\text{limit} &\sim \SI{1}{\mega\eV} \times \left(\frac{\Delta
    \lambda_\text{expt}}{10^{-10}}\right)^{-1 / 2} \times \left(
    \frac{Q_\text{atomic}}{\SI{10}{\eV}}\right)\,, \\
    M^\text{collider}_\text{limit} &\sim \SI{10}{\tera\eV} \times \left(\frac{
    \Delta\lambda_\text{expt}}{10^{-2}}\right)^{-1 / 2} \times \left(
    \frac{Q_\text{collider}}{\SI{1}{\tera\eV}}\right)\,,
\end{align*}
so in both cases, the effective reach of an experiment is extended. Perhaps
surprisingly, the loss in precision in collider experiments is compensated for
by the increase in energy. This is because the variance of our random variable
depends on the resolution length of our experiment.

Because measurements have quantum contributions to their variances, a
direct measurement of this effect may appear challenging. For example,
measuring the variance in the fine-structure constant by observing
the line width of a fine-structure transition would be difficult because of the intrinsic
line width of the transition. Instead, because measurements are in general sensitive
to the square of a matrix element and hence
$\langle \lambda_{\text{expt}}^2 \rangle = \langle \lambda_{\text{expt}} \rangle^2+\Delta\lambda_{\text{expt}}^2$,
the apparent coupling strength will vary with $\Delta\lambda_{\text{expt}}^2= (Q_{\text{expt}} / M_{\ast})^4$.
This leads to an additional source of energy dependence in the observed values of couplings.

\section{Discussion and Future Directions}

A very general feature of string constructions is the appearance of many extra
sectors beyond the visible sector. In this work we have explored one of the
consequences of this visible/extra sector entanglement through the resulting
statistical distribution of couplings.

The main idea pursued in this work is that a helpful way to organize our thinking about the impact of
such extra sectors on the visible sector is in the framework of quantum entanglement. This alone makes it clear
that the visible sector is in general \textit{not} in the ground state, but rather, is in a
mixed state. In general this can lead to a wide variety of possible effects but a model independent and rather generic
feature of this sort of construction is that there is a statistical distribution of couplings. From a practical standpoint,
this suggests that in fitting data to theory one should at least allow this additional variance as an additional measurable
feature.

Having seen that we are really dealing with a mixed state in the visible sector, one might naturally ask whether there are other
observational consequences. In fact, there is a sense in which one implicitly does this whenever one discusses the ``dressed in and out states'' appearing in a scattering amplitude, since there are can be various soft processes that are absorbed into these definitions. It would be interesting to use the present work as a general way to parameterize one's ignorance about asymptotic scattering states.

Our analysis is reminiscent of an old proposal by
Coleman~\cite{Coleman:1988tj}, which argued that an ensemble of wormholes
would also lead to a statistical distribution of physical parameters (for a
recent assessment, see, e.g.,~\cite{Hebecker:2018ofv}). The statistical nature
of couplings considered here is specified over points in spacetime, whereas in
Coleman's case only a single homogeneous value appears. Phenomenologically,
there is no issue with this; it simply reflects the fact that our couplings
really descend from dynamical degrees of freedom. Our result hinges on
entanglement between a visible sector and an extra sector. According to
\cite{Maldacena:2013xja}, such entanglement can perhaps be interpreted as a wormhole
joining geometrically separated regions of a string compactification. In this
sense, the present analysis provides a precise framework for implementing
Coleman's original proposal! Along these lines, it is natural to ask about the
impact of tracing over all sectors (including the Standard Model) other than
those associated with 4D gravity. At low energies, this leads to a statistical
distribution for Newton's constant and the cosmological constant.  Several recent toy models of quantum gravity feature the appearance of a distribution over couplings \cite{Sachdev:1992fk, kitaev, Saad:2019lba, Marolf:2020xie, Balasubramanian:2020jhl} over which the theory averages.  Perhaps these distributions are appearing because all of these theories should really be understood as reduced versions of a complete theory with many unobserved degrees of freedom.

More generally, we can also contemplate the observational consequences of
treating the visible sector as an open system. This would also suggest
potential signatures such as an apparent loss of unitarity and/or CPT
violation. For example, precision fits on the unitarity triangle of the CKM
matrix are on the order of $0.001\%$ to $0.05\%$, and remain quite poorly
constrained for the PMNS matrix~\cite{Agashe:2014kda}. Cosmological variation
in the value of the couplings provides another novel signature~\cite{Davoudiasl:2018ltz}.

All of this points to an exciting new program for probing the stringy origin
of couplings which cuts across several different frontiers of fundamental
physics.

\paragraph{Acknowledgments}
We thank F. Apruzzi, N. Arkani-Hamed, C. Cs\'{a}ki,
M. DeCross, A. Kar, C. Mauger, B. Ovrut, O. Parrikar, R. Penco, C. Rabideau,
M. Reece, J. Ruderman, J. Sakstein, A. Solomon and E. Thomson for helpful
discussions. The work of VB is supported by the Simons Foundation \#385592
through the It From Qubit Simons Collaboration, by the DOE through DE-SC0013528,
and the QuantISED program grant DE-SC0020360. The work of JJH was supported by NSF
CAREER grant PHY-1756996, and is supported by a University of
Pennsylvania University Research Foundation grant and DOE (HEP) Award DE-SC0021484.
The work of EL is supported by DOE grant DE-SC0007901.
The work of APT is supported by DOE (HEP) Award DE-SC0013528.

\appendix

\section{Many Coupled Harmonic Oscillators}
\label{sec:many-osc}

In this Appendix we carry out a more general version of the coupled harmonic oscillator
example given in \cref{sec:osc}. For fields $X(t), Y(t), Z_i(t)$, $i =
1, \dots, N$, consider the 1D Hamiltonian
\begin{equation}
    H = H_X + H_Y + H_Z + H_{X Y} + H_{Y Z}\,,
\end{equation}
with
\begin{equation}
    \begin{aligned}
        H_X &= \frac{P_X^2}{2 m} + \frac{1}{2} m \omega_X^2 X^2\,, \\
        H_Y &= \frac{P_Y^2}{2 M} + \frac{1}{2} M \omega_Y^2 Y^2\,, \\
        H_Z &= \sum_{i = 1}^N \left[\frac{P_{Z_i}^2}{2 M_i} + \frac{1}{2} M_i \omega_i^2 Z_i^2\right]\,, \\
        H_{X Y} &= \varepsilon_{X Y} X Y\,, \\
        H_{Y Z} &= \sum_{i = 1}^N \varepsilon_i Y Z_i\,.
    \end{aligned}
\end{equation}
Here, we think of $X$ as a visible sector field, $Y$ as a modulus, and $Z_i$
as a collection of hidden sector fields. Defining new variables
\begin{equation}
    \begin{pmatrix}x_1  \\ x_2 \\ x_3 \\ \vdots \\ x_{N + 2}\end{pmatrix} = \mu^{-1 / 2} \begin{pmatrix}m^{1 / 2} X \\ M^{1 / 2} Y \\ M_1^{1 / 2} Z_1 \\ \vdots \\ M_N^{1 / 2} Z_N\end{pmatrix}\,, \quad \begin{pmatrix}p_1 \\ p_2 \\ p_3 \\ \vdots \\ p_{N + 2}\end{pmatrix} = \mu^{1 / 2} \begin{pmatrix}m^{-1 / 2} P_X \\ M^{-1 / 2} P_Y \\ M_1^{-1 / 2} P_{Z_1} \\ \vdots \\ M_N^{-1 / 2} P_{Z_N}\end{pmatrix}\,,
\end{equation}
where $\mu = (m M M_1 \dotsm M_N)^{1 / (N + 2)}$, and positive-definite
coupling matrix
\begin{equation}
    \Sigma_{i j} = \begin{pmatrix}\omega_X^2 & \frac{\varepsilon_{X Y}}{2 \sqrt{m M}} & 0 & 0 & \dotsm & 0 \\ \frac{\varepsilon_{X Y}}{2 \sqrt{m M}} & \omega_Y^2 & \frac{\varepsilon_1}{2 \sqrt{M M_1}} & \frac{\varepsilon_2}{2 \sqrt{M M_2}} & \dotsm & \frac{\varepsilon_N}{2 \sqrt{M M_N}} \\ 0 & \frac{\varepsilon_1}{2 \sqrt{M M_1}} & \omega_1^2 & 0 & \dotsm & 0 \\ 0 & \frac{\varepsilon_2}{2 \sqrt{M M_2}} & 0 & \omega_2^2 & \dotsm & 0 \\ \vdots & \vdots & \vdots & \vdots & \ddots & \vdots \\ 0 & \frac{\varepsilon_N}{2 \sqrt{M M_N}} & 0 & 0 & \dotsm & \omega_N^2\end{pmatrix}\,,
\end{equation}
we can rewrite this Hamiltonian in matrix form as
\begin{equation}
    H = \frac{1}{2 \mu} \sum_{i = 1}^{N + 2} p_i^2 + \frac{1}{2} \mu \sum_{i, j = 1}^{N + 2} \Sigma_{i j} x_i x_j\,.
\end{equation}
The matrix $\Sigma_{i j}$ can be diagonalized by an orthogonal matrix $M_{i
j}$,
\begin{equation}
    \Sigma_{i j} = \sum_{k, \ell = 1}^{N + 2} M_{i k} M_{j \ell} D_{k \ell}\,,
\end{equation}
with $D_{i j} = \sigma_i^2 \delta_{i j}$ a diagonal matrix. Defining new
coordinates by $x_i = M_{i j} \tilde{x}_j, p_i = M_{i j} \tilde{p}_j$, we then
have
\begin{equation}
    H = \frac{1}{2 \mu} \sum_{i = 1}^{N + 2} \tilde{p}_i^2 + \frac{1}{2} \mu \sum_{i = 1}^{N + 2} \sigma_i^2 \tilde{x}_i^2\,.
\end{equation}
The ground state of this system of oscillators is
\begin{equation}
\begin{aligned}
    \psi_0(x_i) &= \left(\frac{\mu}{{\pi}}\right)^{(N + 2) / 4} (\sigma_1 \dotsm \sigma_{N + 2})^{1 / 4} \exp\left[-\frac{\mu}{2} \sum_{i = 1}^{N + 2} \sigma_i \tilde{x}_i^2\right] \\
    &= \left(\frac{\mu}{{\pi}}\right)^{(N + 2) / 4} (\det A)^{1 / 4} \exp\left[-\frac{\mu}{2} \sum_{i, j = 1}^{N + 2} A_{i j} x_i x_j\right]\,,
\end{aligned}
\end{equation}
where $A_{i j} = \sum_{k = 1}^{N + 2} \sigma_k M_{i k} M_{j k} = \Sigma_{i
j}^{1 / 2}$, and thus the density matrix is given by
\begin{equation}
    \rho(x_i, x_i') = \psi_0(x_i) \psi_0^*(x_i') = \left(\frac{\mu}{{\pi}}\right)^{(N + 2) / 2} \sqrt{\det A} \exp\left[-\frac{\mu}{2} \sum_{i, j = 1}^{N + 2} A_{i j} (x_i x_j + x_i' x_j')\right]\,.
\end{equation}

Tracing out the hidden sector fields means tracing out $x_i$ for $i > 2$,
which yields
\begin{equation}
\begin{aligned}
    \rho(x_1&, x_1', x_2, x_2') = \frac{\mu}{\pi} \sqrt{\frac{\det A}{\det\tilde{A}}} \\
    &\times \exp\left[-\frac{\mu}{2} A_{1 1} (x_1^2 + x_1'^2)\right] \exp\left[\frac{\mu}{4} (x_1 + x_1')^2 \sum_{i, j = 3}^{N + 2} \left(\tilde{A}^{-1}\right)_{i j} A_{1 i} A_{1 j}\right] \\
    &\times \exp\left[-\frac{\mu}{2} A_{2 2} (x_2^2 + x_2'^2)\right] \exp\left[\frac{\mu}{4} (x_2 + x_2')^2 \sum_{i, j = 3}^{N + 2} \left(\tilde{A}^{-1}\right)_{i j} A_{2 i} A_{2 j}\right] \\
    &\times \exp\left[-\mu A_{1 2} (x_1 x_2 + x_1' x_2')\right] \exp\left[\frac{\mu}{2} (x_1 + x_1') (x_2 + x_2') \sum_{i, j = 3}^{N + 2} \left(\tilde{A}^{-1}\right)_{i j} A_{1 i} A_{2 j}\right]\,,
\end{aligned}
\end{equation}
where $\tilde{A}_{i j}$ is the matrix found by deleting the first two rows and
columns of $A_{i j}$, indexed such that $i, j = 3, \dots, N + 2$.

We can then compute the variance of $x_2$ as
\begin{equation}
    \Var(x_2) = \Tr(x_2^2 \rho) = \frac{1}{2 \mu} \sqrt{\frac{\det A}{\det\tilde{A}}} \frac{B_{1 1}}{\left(B_{1 1} B_{2 2} - B_{1 2}^2\right)^{3 / 2}}\,,
\end{equation}
where
\begin{equation}
    B_{i j} \equiv A_{i j} - \sum_{k, \ell = 3}^{N + 2} \left(\tilde{A}^{-1}\right)_{k \ell} A_{i k} A_{j \ell}\,.
\end{equation}
In the limit of vanishing coupling, this indeed reproduces the expected result $(2 \mu \omega_2)^{-1}$.

\subsection{Weak Coupling}
\label{sec:many-osc-weak}

As an explicit example of the increase in variance from coupling the modulus to a large number of hidden sector oscillators, we consider here the case where the hidden sector oscillators are weakly coupled or very heavy. In this case, the coupling terms can be treated as a perturbation. For this section, we omit the visible sector fields and consider only the modulus and its coupling to hidden sector fields.

Consider a modulus $x_1$ coupled to many hidden sector fields $x_2, \dots, x_N$ via a Hamiltonian of the form
\begin{equation}
	H = H_0 + \epsilon H_1
\end{equation}
with
\begin{equation}
\begin{aligned}
	H_0 &= \frac{1}{2 \mu} \sum_{i = 1}^{N + 2} p_i^2 + \frac{1}{2} \mu \sum_{i = 1}^N \omega_i^2 x_i^2\,, \\
	H_1 &= \frac{1}{2} \mu \sum_{i = 2}^N \varepsilon_i x_1 x_i\,,
\end{aligned}
\end{equation}
and $\epsilon \ll 1$. We have explicitly pulled out the small parameter $\epsilon$ to make the perturbative expansion clear. The energy eigenstates and associated energies of the unperturbed Hamiltonian $H_0$ are
\begin{equation}
\begin{aligned}
	\psi^{(0)}_{k_1, \dots, k_N}(x_1, \dots, x_n) &= \frac{1}{\sqrt{2^{k_1 + \dotsb + k_N} k_1! \dotsm k_N!}} \left(\frac{\mu}{\pi}\right)^{N / 4} (\omega_1 \dotsm \omega_N)^{1 / 4} \\
	&\qquad \times \exp\left(-\frac{1}{2} \mu \sum_{i = 1}^N \omega_i x_i^2\right) H_{k_1}(\sqrt{\mu \omega_1} x_1) \dotsm H_{k_N}(\sqrt{\mu \omega_N} x_N)\,, \\
	E^{(0)}_{k_1, \dots, k_N} &= \frac{N}{2} + \sum_i k_i \omega_i\,,
\end{aligned}
\end{equation}
where $H_i(x)$ are the physicists' Hermite polynomials. Using the notation $\vec{k} = (k_1, \dots, k_N), \vec{0} = (0, \dots, 0)$, the ground state of the Hamiltonian $H$ takes the form
\begin{equation}
\begin{aligned}
	\ket{\vec{0}} &= \ket{\vec{0}^{(0)}} + \epsilon \ket{\vec{k}} \frac{\braopket{\vec{k}}{H_1}{\vec{0}}}{E_{\vec{0}}^{(0)} - E_{\vec{k}}^{(0)}} \\
	&\qquad + \epsilon^2 \left[\ket{\vec{k}} \left(\frac{\braopket{\vec{k}}{H_1}{\vec{\ell}} \braopket{\vec{\ell}}{H_1}{\vec{0}}}{\left(E_{\vec{0}}^{(0)} - E_{\vec{k}}^{(0)}\right) \left(E_{\vec{0}}^{(0)} - E_{\vec{\ell}}^{(0)}\right)} - \frac{\braopket{\vec{k}}{H_1}{\vec{0}} \braopket{\vec{0}}{H_1}{\vec{0}}}{\left(E_{\vec{0}}^{(0)} - E_{\vec{k}}^{(0)}\right)^2}\right)\right. \\
	&\qquad\qquad\qquad \left.- \ket{\vec{0}} \frac{\abs*{\braopket{\vec{k}}{H_1}{\vec{0}}}^2}{2 \left(E_{\vec{0}}^{(0)} - E_{\vec{k}}^{(0)}\right)^2}\right] \\
	&\qquad + O\left(\epsilon^3\right)\,,
\end{aligned}
\end{equation}
where in each term there is implicit summation over all nonzero values of the vectors $\vec{k}, \vec{\ell}$ that appear. Rewriting the perturbing Hamiltonian in terms of creation and annihilation operators,
\begin{equation}
	H_1 = \sum_{i = 2}^N \frac{\varepsilon_i}{4 \sqrt{\omega_1 \omega_i}} \left(a_1 + a_1^\dag\right) \left(a_i + a_i^\dag\right)\,,
\end{equation}
we see that the only relevant nonzero matrix elements are
\begin{equation}
\begin{aligned}
	\braopket{1_1 i_1}{H_1}{\vec{0}} &= \frac{\varepsilon_i}{4 \sqrt{\omega_1 \omega_i}}\,, \\
	\braopket{1_2 i_2}{H_1}{1_1 i_1} &= \frac{\varepsilon_i}{2 \sqrt{\omega_1 \omega_i}}\,, \\
	\braopket{1_2 i_1 j_1}{H_1}{1_1 i_1} &= \frac{\varepsilon_j}{2 \sqrt{2 \omega_1 \omega_j}}\,, \\
	\braopket{1_2}{H_1}{1_1 i_1} &= \frac{\varepsilon_i}{2 \sqrt{2 \omega_1 \omega_i}}\,, \\
	\braopket{i_2}{H_1}{1_1 i_1} &= \frac{\varepsilon_i}{2 \sqrt{2 \omega_1 \omega_i}}\,, \\
	\braopket{i_1 j_1}{H_1}{1_1 i_1} &= \frac{\varepsilon_j}{4 \sqrt{\omega_1 \omega_j}}\,,
\end{aligned}
\end{equation}
where we are using the shorthand, e.g., $\ket{1_1 i_1} = \ket{1,0,\dots,0,\underbrace{1}_{k_i},0,\dots,0}$. We see then that
\begin{equation}
\begin{aligned}
	\ket{\vec{0}} &= \ket{\vec{0}^{(0)}} - \epsilon \frac{\varepsilon_i}{4 \sqrt{\omega_1 \omega_i} (\omega_1 + \omega_i)} \ket{1_1 i_1} \\
	&\qquad + \epsilon^2 \left[\frac{\varepsilon_i^2}{16 \omega_1 \omega_i (\omega_1 + \omega_i)^2} \ket{1_2 i_2} + \frac{\varepsilon_i \varepsilon_j}{8 \sqrt{2 \omega_i \omega_j} \omega_1 (2 \omega_1 + \omega_i + \omega_j) (\omega_1 + \omega_i)} \ket{1_2 i_1 j_1}\right. \\
	&\qquad\qquad\qquad + \frac{\varepsilon_i^2}{16 \sqrt{2} \omega_1^2 \omega_i (\omega_1 + \omega_i)} \ket{1_2} + \frac{\varepsilon_i^2}{16 \sqrt{2} \omega_1 \omega_i^2 (\omega_1 + \omega_i)} \ket{i_2} \\
	&\qquad\qquad\qquad \left.+ \frac{\varepsilon_i \varepsilon_j}{16 \sqrt{\omega_i \omega_j} \omega_1 (\omega_i + \omega_j) (\omega_1 + \omega_i)} \ket{i_1 j_1} - \frac{\varepsilon_i^2}{32 \omega_1 \omega_i (\omega_1 + \omega_i)^2} \ket{\vec{0}}\right] \\
	&\qquad + O\left(\epsilon^3\right)\,,
\end{aligned}
\end{equation}
where we are implicitly summing over the indices $i, j = 2, \dots, N$. The wavefunction is thus given by
\begin{equation}
\begin{aligned}
	\psi_{\vec{0}}(\vec{x}) &= \left(\frac{\mu}{\pi}\right)^{N / 4} (\omega_1 \dotsm \omega_N)^{1 / 4} \exp\left(-\frac{1}{2} \mu \sum_{i = 1}^N \omega_i x_i^2\right) \\
	&\qquad \times \Bigg\{1 - \epsilon \frac{\varepsilon_i H_1(\sqrt{\mu \omega_1} x_1) H_1(\sqrt{\mu \omega_i} x_i)}{8 \sqrt{\omega_1 \omega_i} (\omega_1 + \omega_i)} \\
	&\qquad\qquad\quad + \epsilon^2 \left[\frac{\varepsilon_i^2 H_2(\sqrt{\mu \omega_1} x_1) H_2(\sqrt{\mu \omega_i} x_i)}{128 \omega_1 \omega_i (\omega_1 + \omega_i)^2}\right. \\
	&\qquad\qquad\qquad\qquad\quad + \frac{\varepsilon_i \varepsilon_j H_2(\sqrt{\mu \omega_1} x_1) H_1(\sqrt{\mu \omega_i} x_i) H_1(\sqrt{\mu \omega_j} x_j)}{64 \sqrt{\omega_i \omega_j} \omega_1 (2 \omega_1 + \omega_i + \omega_j) (\omega_1 + \omega_i)} \\
	&\qquad\qquad\qquad\qquad\quad + \frac{\varepsilon_i^2 H_2(\sqrt{\mu \omega_1} x_1)}{64 \omega_1^2 \omega_i (\omega_1 + \omega_i)} + \frac{\varepsilon_i^2 H_2(\sqrt{\mu \omega_i} x_i)}{64 \omega_1 \omega_i^2 (\omega_1 + \omega_i)} \\
	&\qquad\qquad\qquad\qquad\quad \left.+ \frac{\varepsilon_i \varepsilon_j H_1(\sqrt{\mu \omega_i} x_i) H_1(\sqrt{\mu \omega_j} x_j)}{32 \sqrt{\omega_i \omega_j} \omega_1 (\omega_i + \omega_j) (\omega_1 + \omega_i)} - \frac{\varepsilon_i^2}{32 \omega_1 \omega_i (\omega_1 + \omega_i)^2}\right] \\
	&\qquad\qquad\quad + O\left(\epsilon^3\right)\Bigg\}\,.
\end{aligned}
\end{equation}
From this, we can read off the reduced density matrix for the modulus $x_1$:
\begin{equation}
\begin{aligned}
	\rho(x_1, x_1') &= \psi_{\vec{0}}(\vec{x}) \psi_{\vec{0}}^*(\vec{x}') \\
	&= \sqrt{\frac{\mu \omega_1}{\pi}} \exp\left[-\frac{1}{2} \mu \omega_1 \left(x_1^2 + x_1'^2\right)\right] \\
	&\qquad \times \Bigg[1 + \epsilon^2 \frac{\varepsilon_i^2 \left(H_2(\sqrt{\mu \omega_1} x_1) + H_2(\sqrt{\mu \omega_1} x_1')\right)}{64 \omega_1^2 \omega_i (\omega_1 + \omega_i)} \\
	&\qquad\qquad\qquad + \epsilon^2 \frac{\varepsilon_i^2 \left(H_1(\sqrt{\mu \omega_1} x_1) H_1(\sqrt{\mu \omega_1} x_1') - 2\right)}{32 \omega_1 \omega_i (\omega_1 + \omega_i)^2} + O\left(\epsilon^4\right)\Bigg] \\
	&= \sqrt{\frac{\mu \omega_1}{\pi}} \exp\left[-\frac{1}{2} \mu \omega_1 \left(x_1^2 + x_1'^2\right)\right] \\
	&\qquad \times \Bigg[1 + \epsilon^2 \frac{\varepsilon_i^2 \left(\mu \omega_1 x_1^2 + \mu \omega_1 x_1'^2 - 1\right)}{16 \omega_1^2 \omega_i (\omega_1 + \omega_i)} + \epsilon^2 \frac{\varepsilon_i^2 \left(2 \mu \omega_1 x_1 x_1' - 1\right)}{16 \omega_1 \omega_i (\omega_1 + \omega_i)^2} + O\left(\epsilon^4\right)\Bigg]\,.
\end{aligned}
\end{equation}
Returning to explicit summation over indices, we thus find that the variance of $x_1$ is
\begin{equation}
	\Var(x_1) = \Tr\left(x_1^2 \rho\right) = \frac{1}{2 \mu \omega_1} + \sum_{i = 2}^N \frac{\epsilon^2 \varepsilon_i^2 (2 \omega_1 + \omega_i)}{16 \mu \omega_1^3 \omega_i (\omega_1 + \omega_i)^2} + O\left(\epsilon^4\right)\,.
\end{equation}
We see that, as expected, the variance goes to the usual value $(2 \mu \omega_1)^{-1}$ in the limit of zero coupling, and grows quadratically with the coupling.


\bibliographystyle{utphys}
\bibliography{references}

\end{document}